\documentstyle[11pt,epsf]{article}
\title{
  {\vspace{-3cm} \normalsize \hfill
    \parbox{38mm}{HD-THEP-99-9 \\
                  cond-mat/9902354}}\\[25mm]
A two step algorithm for learning from
unspecific reinforcement  }

\author{
Reimer K\"uhn$^1$,
Ion-Olimpiu Stamatescu$^{2,3}$\\[5mm]
{\small $^1$Institut f\"ur Theoretische Physik, Universit\"at Heidelberg,
 Philosophenweg 19}\\
{\small D-69120, Heidelberg, Germany}\\
{\small $^2$Institut f\"ur Theoretische Physik, Universit\"at Heidelberg,
 Philosophenweg 16}\\
{\small D-69120, Heidelberg, Germany}\\
{\small $^3$FESt, Schmeilweg 5, D-69118, Heidelberg, Germany}
}

\date{February 26, 1999}

\newcommand{\be}{\begin{equation}}
\newcommand{\ee}{\end{equation}}
\newcommand{\bea}{\begin{eqnarray}}
\newcommand{\eea}{\end{eqnarray}}
\newcommand{\no}{\noindent}
\newcommand{\sign}{{\rm sign}}

\begin{document}
\maketitle

\begin{abstract}
We study a simple learning model based on the Hebb rule to cope with 
``delayed", unspecific reinforcement. In spite of the unspecific nature 
of the information-feedback, convergence to asymptotically perfect 
generalization is observed, with a rate depending, however, in a 
non-universal way on learning parameters. Asymptotic convergence can 
be as fast as that of Hebbian learning, but may be slower. Morever, for
a certain range of parameter settings, it depends on initial conditions
whether the system can reach the regime of asymptotically perfect 
generalization, or rather approaches a stationary state of poor 
generalization.
 
\end{abstract}

\section{Introduction}

 Introducing biologically motivated features in
models for learning has usually a double role: testing hypotheses  
for natural learning and finding hints for artificial learning. 
These problems can be stated at various sophistication levels. 
Here we  
do not take the more ambitious point of view of describing  
the complexity of the former or of finding  
optimal  
algorithms for the latter. 
On the contrary, our motivation is to investigate which are the  
capabilities of 
very  elementary  
mechanisms.  
 
One urgent problem with which a system, either natural or artificial, 
may be  confronted when 
trying to improve its performance is to learn only from the final  
success/failure of 
{\it series} of consecutive decisions.  
The typical situation we may consider is that of an ``agent" which  
let free in a complicated ``landscape" tries many ``paths" to reach a ``goal" 
and has to optimize its path (a local problem) knowing only the  
``time" (or cost) it needs to reach the goal (global information). Here  
``goal" may be a survival interest or the solution of a problem, 
``path" a series of moves or of partial solution steps in a 
complex geographical or mathematical ``landscape"  
etc.  
The problem we want to approach here is to find out whether  
there are elementary features characterizing learning 
under such {\it unspecific} reinforcement conditions.  
From the point of view of reinforcement learning our problem may be  
seen under the 
``class III" problems in the classification of Hertz et al. \cite{her}.    
However, we stress that our attitude is not that of finding good algorithms  
for tackling  
special problems, like movement, control or games -- see, e.g., \cite{kael}. 
 For this reason we  
do not consider evolved algorithms from the class of 
 Q-learning \cite{wat}, of TD learning \cite{sut}, agent and  
critic \cite{bart}, etc but restrict to most primitive 
 algorithms which we may think of having a chance to have developed  
under natural conditions. On the other hand, if such algorithms will 
prove capable of tackling the problem they may well give further 
insights.\footnote{An illustration of the problem was provided in an early paper \cite{MS85} 
dealing with  
these questions in the simulation of a device 
moving on a board.} 
 
In the case of 
 neural network systems the normal situation is already that of lacking 
detailed control over the synapses and learning is achieved by  
confronting the ``pupil" system with the correct answer after each 
presentation of a pattern. For perceptrons both  the unsupervised 
Hebb rule and the supervised perceptron algorithm are known to lead 
to asymptotically perfect generalization, although with different 
asymptotic laws. 
In our problem setting, however, the pupil never knows the right answer  
to each question, but only the average error it makes over many tests.  
In previous work concerned with this problem  
\cite{mai98} (see also \cite{old}) we presented  
an analysis of a 2-step algorithm based on the Hebb rule 
for perceptrons and used 
computer simulations and a rough approximation 
to estimate the convergence conditions. In the present work 
we undertake a detailed study of this learning algorithm 
 which we  call for simplicity  
``association-reinforcement(AR)-Hebb-rule". This algorithm introduces two 
learning parameters and we find that its generalization behaviour is
highly nontrivial: in the pre-asymptotic region  and depending on the
network parameters {\em fixed points\/} of the learning dynamics may appear. 
This leads either to asymptotically perfect generalization with non-universal 
power laws depending on the (ratio of the) learning parameters,
or to stationary states of very poor generalizationi, according to the
network parameters and initial conditions.
 
That this AR-Hebb-algorithm may be of a more general interest 
is suggested by applying it to a concrete problem of optimizing paths 
in a landscape with obstacles and traps, in a neural network recasting of  
\cite{MS85}; this study will be  
presented elsewhere (partial results have been given in \cite{mai98}).  
 
In the next section we shall introduce the problem and the algorithm,  
and in section 3 we shall present results from numerical simulations. 
In section 4 we shall study a coarse grained approximation which  
is appropriate for large networks (``thermodynamic limit"). Section 5 is  
reserved for conclusions.

\section{Learning rule for perceptrons under unspecific reinforcement} 
 
We consider perceptrons with Ising units $s,s_i = \pm 1$  
and real weights (synapses) $J_i$: 
\begin{equation} 
     s = \sign \left( \frac{1}{\sqrt{N}}\sum_{i=1}^N J_i s_i\right) =  
         \sign \left( \frac{1}{\sqrt{N}}{\bf J}\cdot{\bf {s}}\right) 
\label{e.1} 
\end{equation} 
\no Here $N$ is the number of input nodes, and we put no explicit thresholds.  
The network (pupil) is presented with a series of patterns  
$\xi_i^{(q,l)}$,  $q\in {\rm I\!N}$, $l=1, ... ,L$ 
to which it answers with $s^{(q,l)}$. A training period consists of 
the successive presentation of $L$ patterns. 
The answers are compared with the corresponding answers  $t^{(q,l)}$ 
of a teacher 
with pre-given weights $B_i$ and 
the average error made by the pupil over one 
training period is calculated: 
\be 
    e_q = \frac{1}{2L} \sum_{l=1}^L |t^{(q,l)} -s^{(q,l)}| . 
\label{e.2} 
\ee 
\no The training algorithm 
consists of two parts:\par 
\begin{itemize} 
\item [I.] - a ``blind" Hebb-type  {\it association} at each presentation of a pattern: 
\be 
   {\bf J}^{(q,l+1)} ={\bf J}^{(q,l)} + \frac{a_1}{\sqrt{N}} s^{(q,l)}
{\bf \xi}^{(q,l)}; 
\label{e.3} 
\ee 
\item [II.] - an ``unspecific" but graded {\it reinforcement} proportional
to the average error $e_q$ introduced in (\ref{e.2}), also Hebbian, 
at the end of each training period, 
\be 
   {\bf J}^{(q+1,1)} = {\bf J}^{(q,L+1)} - \frac{a_2}{\sqrt{N}} e_q 
                \sum_{l=1}^L s^{(q,l)}{\bf \xi}^{(q,l)}. 
\label{e.4} 
\ee 
\end{itemize} 
\no Because of these 2 steps we  call this algorithm  
 ``association/reinforcement(AR)-Hebb-rule" (or ``2-Hebb-rule",\cite{mai98}).  
We are  
interested in the behavior with the number of iterations $q$ of the generalization error $\epsilon_g(q)$: 
 
\be 
   \epsilon_g(q) = \frac{1}{\pi}{\rm arccos} 
\left({{{\bf J}\cdot {\bf B}} \over {|{\bf J}|~|{\bf B}|}}\right) 
\label{e.errg} 
\ee 
 
\no The training patterns ${\bf \xi}^{(q,l)}$ are generated randomly, and are 
taken to be unbiased in the present paper. The case of structured patterns is 
more complicated, and will be dealt with in a separate publication \cite{bi+99}. We shall test whether the behavior of $\epsilon_g(q)$ follows a power law at 
large $q$: 
\be 
      \epsilon_g(q) \simeq {\rm const}~ q^{-p} 
\label{e.6} 
\ee

\no Notice the following features: 
\begin{itemize} 
\item [a)] During training the pupil only uses its own associations 
${\bf \xi}^{(q,l)} 
\leftrightarrow s^{(q,l)}$ and the average error $e_q$ which does not refer specifically
 to the particular steps $l$. 
\item [b)] Since the answers $s^{(q,l)}$ are made on the basis of the instantaneous 
weight values {\bf J}$^{(q,l)}$ which change at each step according to eq. (3), the series of  
answers form a correlated sequence with each step depending on the previous one. 
Therefore $e_q$ measures in fact the performance of a ``path", an interdependent set of
 decisions. 
\item [c)] For $L=1$ the algorithm reproduces the usual  
``perceptron rule" (for  
$a_1=0$) or to the usual ``unsupervised Hebb rule" (for $a_2 = 2 a_1$)
for on-line learning, for which the corresponding asymptotic behavior
is known\cite{biri}, \cite{val}. 
\end{itemize}

\section{Numerical results} 
 
In a preliminary analysis \cite{mai98} we have tested various combinations 
of $L =1, 5,10,15$ and $N=50,100,200,300$.  We went with  
 $q$ up to $4.10^5$. We found  the convergence of the learning procedure  
to depend on the ratio $a_1/a_2$, in particular  no convergence was  
found for $L$ of 5 and higher  
if this ratio was decreased significantly below $0.2$. For fixed $a_1, a_2$ 
the asymptotic behavior with $q$ appeared well 
reproduced by a power law and {\it the exponent was found to depend on $L$}. 
For $L=1$ varying $a_1/a_2$ between 0 and 1/2  
interpolates between perceptron and Hebbian learning, for ratios larger than 1
new asymptotic behavior can be expected to show up (see sect. 4)  --
we did not perform a systematic numerical analysis for $L=1$, however.

\begin{figure}[htb] 
\vspace{12cm} 
\includegraphics{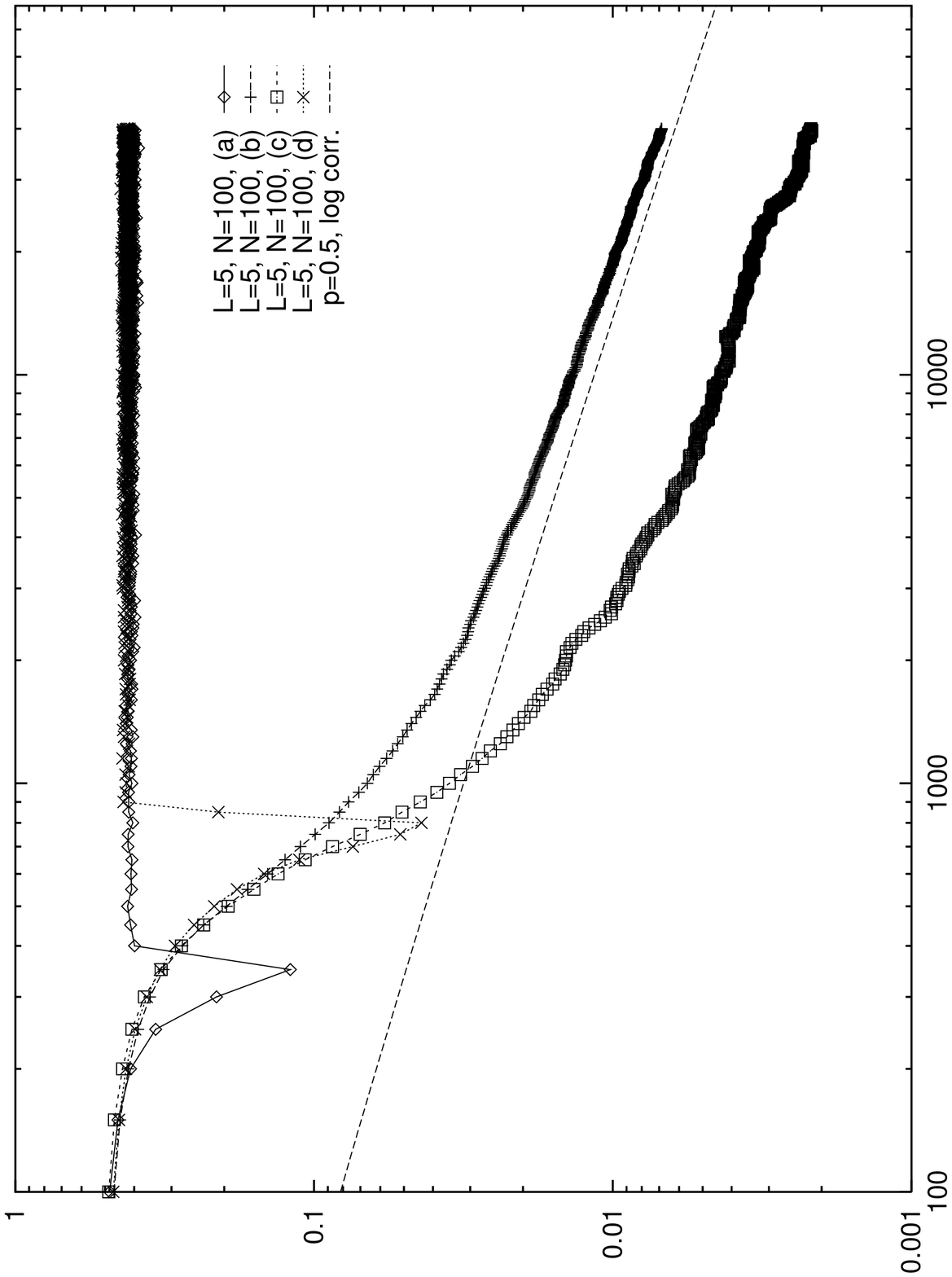} 
\includegraphics{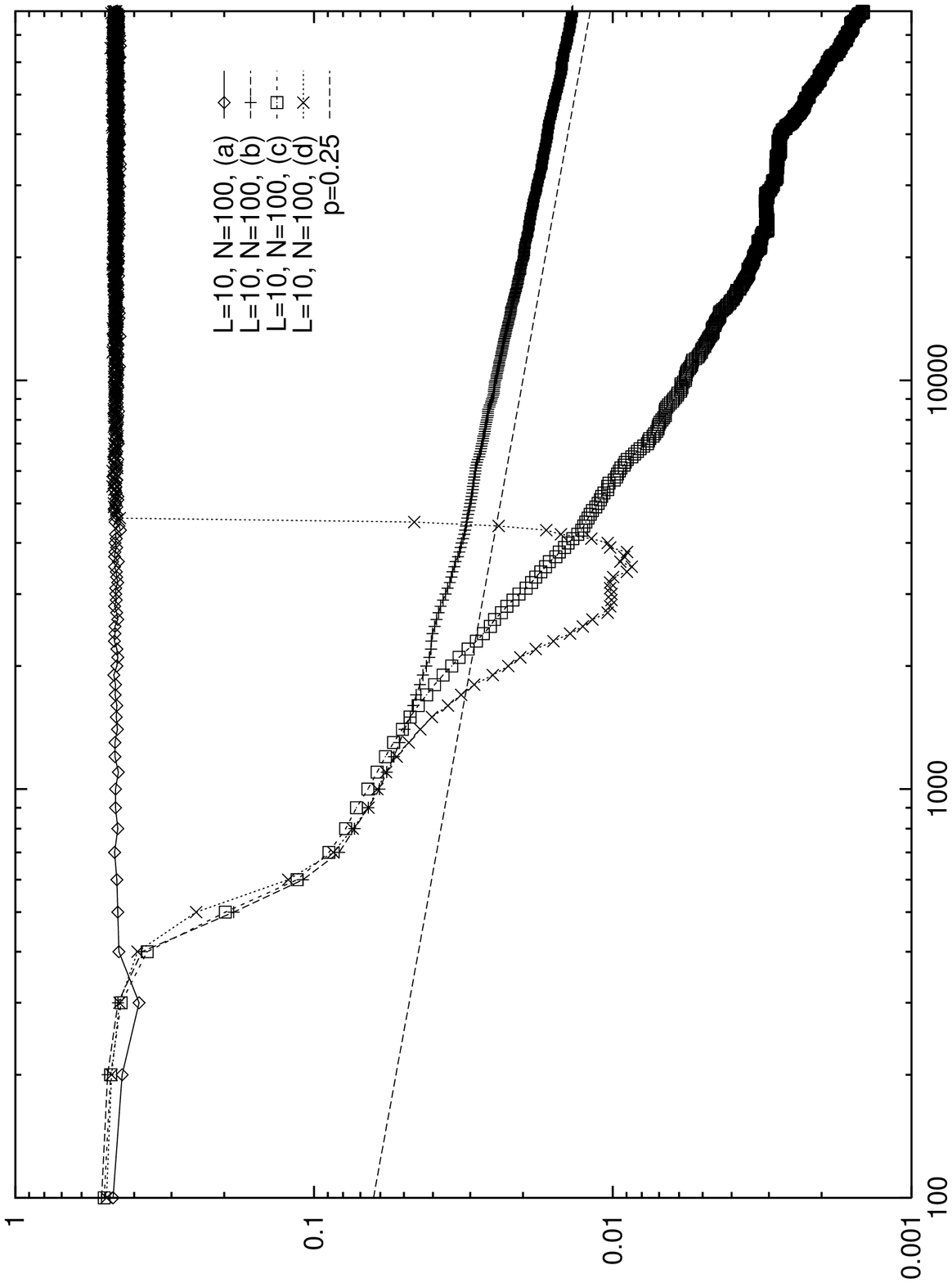} 
\includegraphics{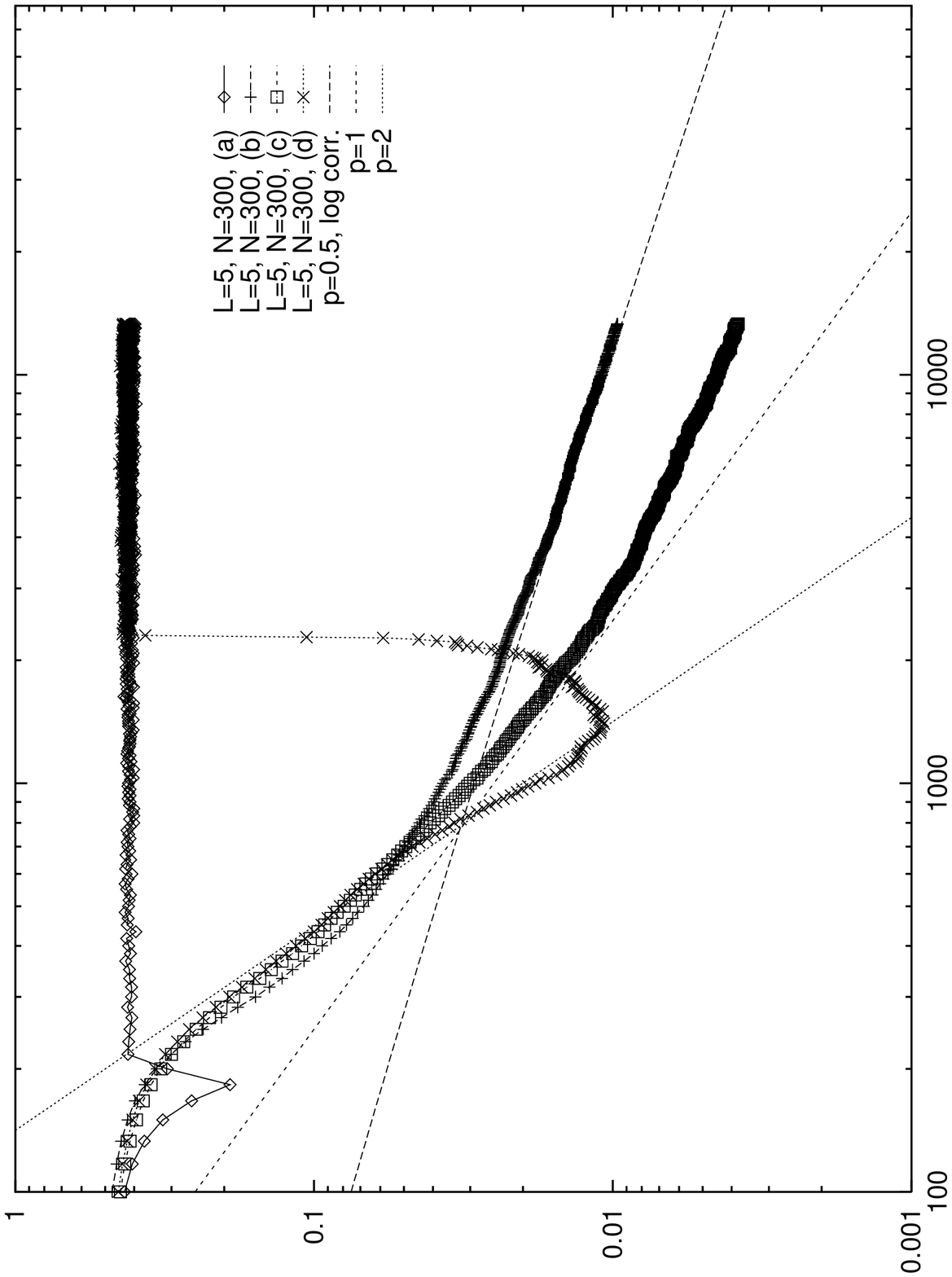} 
\includegraphics{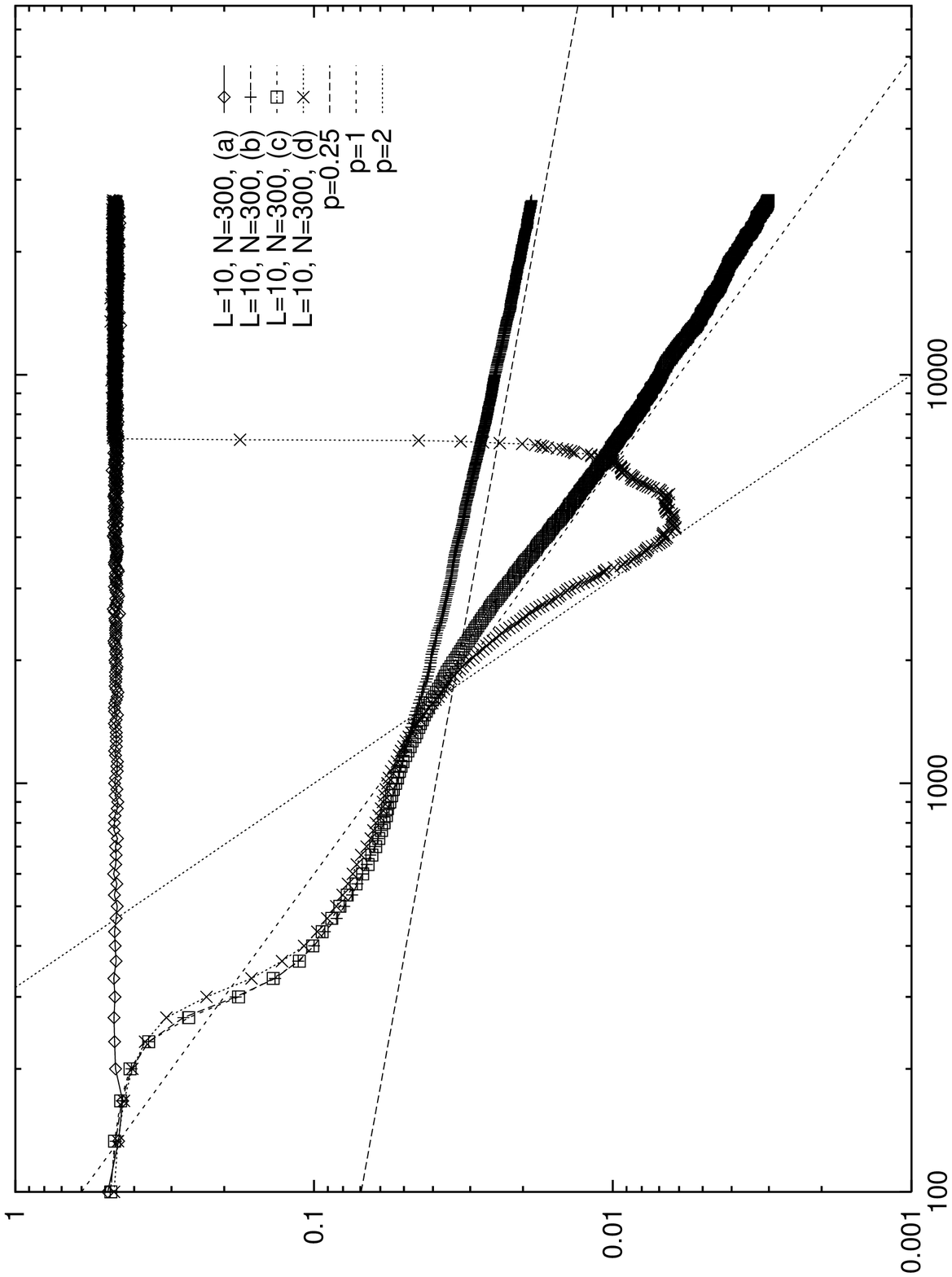} 
\caption{{\it Generalization error $\epsilon_g$ vs. $\alpha$ for $N = 100$,  
$L=5$ and $L=10$ (upper plots) and $N=300$, $L=5$ and $L=10$ (lower plots),  
for the algorithms (a - d) of eqs. (\ref{e.aa}-\ref{e.ad}). The lines  
indicate the expected asymptotic behaviour as suggested by the coarse  
grained approximation discussed in section 4 for the corresponding  
$a_1/a_2$ ratio, as well as 2 further power laws for illustration}. 
} 
\label{f.sim} 
\end{figure}

In the present, more precise analysis we use  
$L=5,10$ and $N=100,300$, going up to $8.10^5$ iterations.  
We introduce: 
 
\be 
\alpha=qL/N . 
\ee 
 
\no We present here results for the following 
choices of parameters: 
\bea 
a_2=0.012, &&{\rm and}\\ 
(a)\ \ &a_1&= a_2/20; \label{e.aa}\\ 
(b)\ \ &a_1&= a_2/5; \label{e.ab} \\ 
(c)\ \ &a_1&= a_2/5  \ {\rm for}\ \alpha < 100L, \nonumber \\  
 &a_1&=a_2/(2L)  \ {\rm for}\ \alpha \geq 100L; \label{e.ac}\\ 
(d)\ \ &a_1&= a_2/5  \ {\rm for}\ \alpha < 100L,\nonumber \\  
 &a_1&=0.  \ {\rm for}\ \alpha \geq 100L. \label{e.ad} 
\eea 
 
\no We use random initial conditions with the same normalization for  
the teacher and pupil weights, ${\bf B}^2/N = {\bf J}^2/N = 1$.  
The results are shown in Fig. \ref{f.sim}.  
In agreement with the preliminary  
results of \cite{mai98} we find no convergence in the case $(a)$  
and convergence in the case $(b)$.  If 
a certain threshold  in $\epsilon_g$ is achieved, switching to 
a smaller ratio $a_1/a_2$ is seen to accelerate the asymptotic convergence
-- case $(c)$ --, 
but even then $a_1$ cannot be set to zero -- case $(d)$. Similar behaviour
is observed for other $N$ and $L \ge 5$.
 
This intriguing behaviour incited us to try to obtain analytic  
understanding by using the coarse grained analysis discussed in the next 
section.

\section{Coarse grained analysis} 
 
We combine {\it blind association} (\ref{e.3}) during a learning period of $L$ 
elementary steps and the graded {\it unspecific reinforcement} (\ref{e.4}) at 
the end of each learning period into one coarse grained step 
 
\bea  
{\bf J}^{(q+1,1)} = {\bf J}^{(q,1)} + \frac{1}{\sqrt{N}}(a_1-a_2 e_q)\sum_{l=1}^L 
\sign({\bf J}^{(q,l)}\cdot{\bf \xi}^{(q,l)})\ {\bf \xi}^{(q,l)}, \label{e.cgu}\\ 
e_q = \frac{1}{2L}\sum_{l=1}^L|\sign({\bf B}\cdot{\bf \xi}^{(q,l)}) -  
\sign({\bf J}^{(q,l)}\cdot{\bf \xi}^{(q,l)})|.\label{e.cge} 
\eea 
 
\no We introduce the notations: 
 
\be 
{\hat R}(q,l) = \frac{1}{N}{\bf B}\cdot{\bf J}^{(q,l)}\quad ,\quad 
 {\hat Q}(q,l) = \frac{1}{N}[{\bf J}^{(q,l)}]^2 
 \label{e.cgdef} 
 \ee 
 
\no and we  normalize the teacher weights to 1, i.e. ${\bf B}^2/N = 1$ . 
In the ``thermodynamic limit" $L/N \rightarrow 0$ one can treat $\alpha$ 
as a continuous variable. We shall follow standard procedures \cite{her}, 
\cite{val}, \cite{kica}, \cite{bisch} and obtain the following expressions for 
the changes of ${\hat R}$ and ${\hat Q}$ over a coarse grained step: 
 
\bea 
L\frac{d}{d\alpha}{\hat R} &=&  \frac{1}{\sqrt{N}}(a_1-a_2 e_q)\sum_{l=1}^L 
\sign({\bf J}^{(q,l)}\cdot{\bf \xi}^{(q,l)})({\bf B}\cdot{\bf \xi}^{(q,l)}), 
\label{e.cgd1}\\ 
L\frac{d}{d\alpha}{\hat Q} &=& \frac{2}{\sqrt{N}}(a_1-a_2 e_q)\sum_{l=1}^L \sign({\bf J}^{(q,l)}\cdot{\bf \xi}^{(q,l)})({\bf J}^{(q,l)}\cdot{\bf \xi}^{(q,l)}) 
\nonumber\\ 
&+&\frac{1}{N}(a_1-a_2 e_q)^2\left(\sum_{l=1}^L 
\sign({\bf J}^{(q,l)}\cdot{\bf \xi}^{(q,l)})\ {\bf \xi}^{(q,l)}\right)^2. \label{e.cgd2} 
\eea 
 
\no In the following we shall consider unbiased random input--patterns with
 
\be 
\langle \xi_i^{(l,q)}\xi_j^{(k,r)}\rangle = \delta_{ij}\delta_{lk}\delta_{qr}. 
\ee 
 
\no The local fields: 
 
\be h_J^{(q,l)} = \frac{1}{\sqrt{N}}{\bf J}^{(q,l)}\cdot{\bf \xi}^{(q,l)},\  
 h_B^{(q,l)} = \frac{1}{\sqrt{N}}{\bf B}\cdot{\bf \xi}^{(q,l)}, \label{e.cgh} 
\ee 
 
\no are then normally distributed with second moments  
 
\be  
\langle (h_J^{(q,l)})^2 \rangle = \langle {\hat Q}(q,l)\rangle = Q\quad ,\quad  
\langle (h_B^{(q,l)})^2 \rangle = 1\quad ,\quad 
\langle (h_J^{(q,l)}h_B^{(q,l)}) \rangle = \langle {\hat R(q,l)} \rangle  = R. 
\label{e.cgf} 
\ee 
 
\no Their joint probability density is thus given by 

\bea 
p(h_J,h_B) &=& \frac{1}{2\pi \sqrt{\Delta}} {\rm exp}\left(  
-\frac{1}{2\Delta}(Q h_B^2 -2Rh_Jh_B+h_J^2)\right),\label{e.cgm}
\eea
with
\bea 
\Delta&=& Q-R^2\ . \label{e.del} 
\eea 
 
\no In the thermodynamic limit $N\to \infty$, the self-overlap of the 
learner ${\hat Q} $ and its overlap $\hat R$ with the teacher are self--averaging, so that their evolution equations (\ref{e.cgd1}),
(\ref{e.cgd2}) can be directly rewritten in terms of evolution equations 
for their averages. Moreover, these averages $Q$ and $R$ become smooth 
functions on the $\alpha$--scale, so that we can neglect the dependence of $R$ 
and $Q$ on $l$ in (\ref{e.cgm}) when used to perform averages on the right 
hand sides of (\ref{e.cgd1}) and (\ref{e.cgd2}), as it would only produce 
${\cal O}(1/N)$ corrections to the evolution equations, which become negligible as $N\to\infty$. One thus obtains
 
\bea 
\frac{dR}{d\alpha} &=& \sqrt{\frac{2}{\pi}}  \left[ a_1\frac{R}{\sqrt{Q}} 
- \frac{a_2}{2}\left( \frac{R}{\sqrt{Q}} -\frac{1}{L} - (1-\frac{1}{L})  
P \frac{R}{\sqrt{Q}}\right) \right] \label{e.dr} \\ 
\frac{dQ}{d\alpha} &=& 2 \sqrt{\frac{2}{\pi}}  \left[ a_1\sqrt{Q} - 
 \frac{a_2}{2}\left( \sqrt{Q} -  \frac{R}{L} - (1-\frac{1}{L})P\sqrt{Q}\right) 
\right]\nonumber\\ 
&+& \left[ a_1^2  - a_1a_2 (1-P) + \frac{a_2^2}{4} \left( 1 -2P +\frac{1}{L} + 
(1-\frac{1}{L})P^2\right) \right]\ , \label{e.dq2} 
\eea 
 
\no where 
 
\bea  
P &=& - \frac{1}{\pi\sqrt{\Delta}}\int_{-\infty}^{\infty} 
\hbox{e}^{-\frac{1}{2}x^2}\hbox{sign}(x)\int_{Rx}^{\infty}dy 
\hbox{e}^{-\frac{1}{2\Delta}y^2} \nonumber \\ 
&=& 1-\frac{2}{\pi}\hbox{arccos}\left(\frac{R}{\sqrt{Q}}\right). \label{e.P}  
\eea 
 
\no The generalization error is: 
 
\be  
\epsilon_g = \frac{1}{\pi}{\rm arccos}\left(\frac{R}{\sqrt{Q}}\right)\ . \label{e.x2}
\ee 
 
\no We may formally eliminate one of the learning parameters by rescaling our quantities by the parameter $a_2$: 
 
\be  
R = {\cal R}a_2,\ \ Q = {\cal Q}a_2^2, \ \ \lambda =\frac{a_1}{a_2}, 
\label{e.eno}
\ee 
 
\no We then obtain: 
 
\bea 
\frac{d\epsilon_g}{d\alpha}&=& -\frac{1}{\sqrt{2\pi}\pi L \sqrt{\cal Q}} 
{\rm sin}(\pi \epsilon_g)  \nonumber \\ 
&+&\frac{1}{2\pi {\cal Q}}{\rm cotg}(\pi \epsilon_g) 
\left(\lambda^2 -(2\lambda - \frac{1}{L})\epsilon_g + (1-\frac{1}{L}) 
\epsilon_g^2\right), \label{e.deps}\\ 
\frac{d \sqrt{\cal Q}}{d\alpha} &=&\sqrt{\frac{2}{\pi}}\left(\lambda - 
(1-\frac{1}{L})\epsilon_g - \frac{1}{2L}\Big(1-{\rm cos}(\pi 
\epsilon_g)\Big)\right) 
 \nonumber \\ 
&+&\frac{1}{2\sqrt{\cal Q}}\left(\lambda^2 -(2\lambda - \frac{1}{L})\epsilon_g + 
(1-\frac{1}{L})\epsilon_g^2\right). \label{e.dqel} 
\eea

\begin{figure}[htb] 
\vspace{7cm} 
\includegraphics{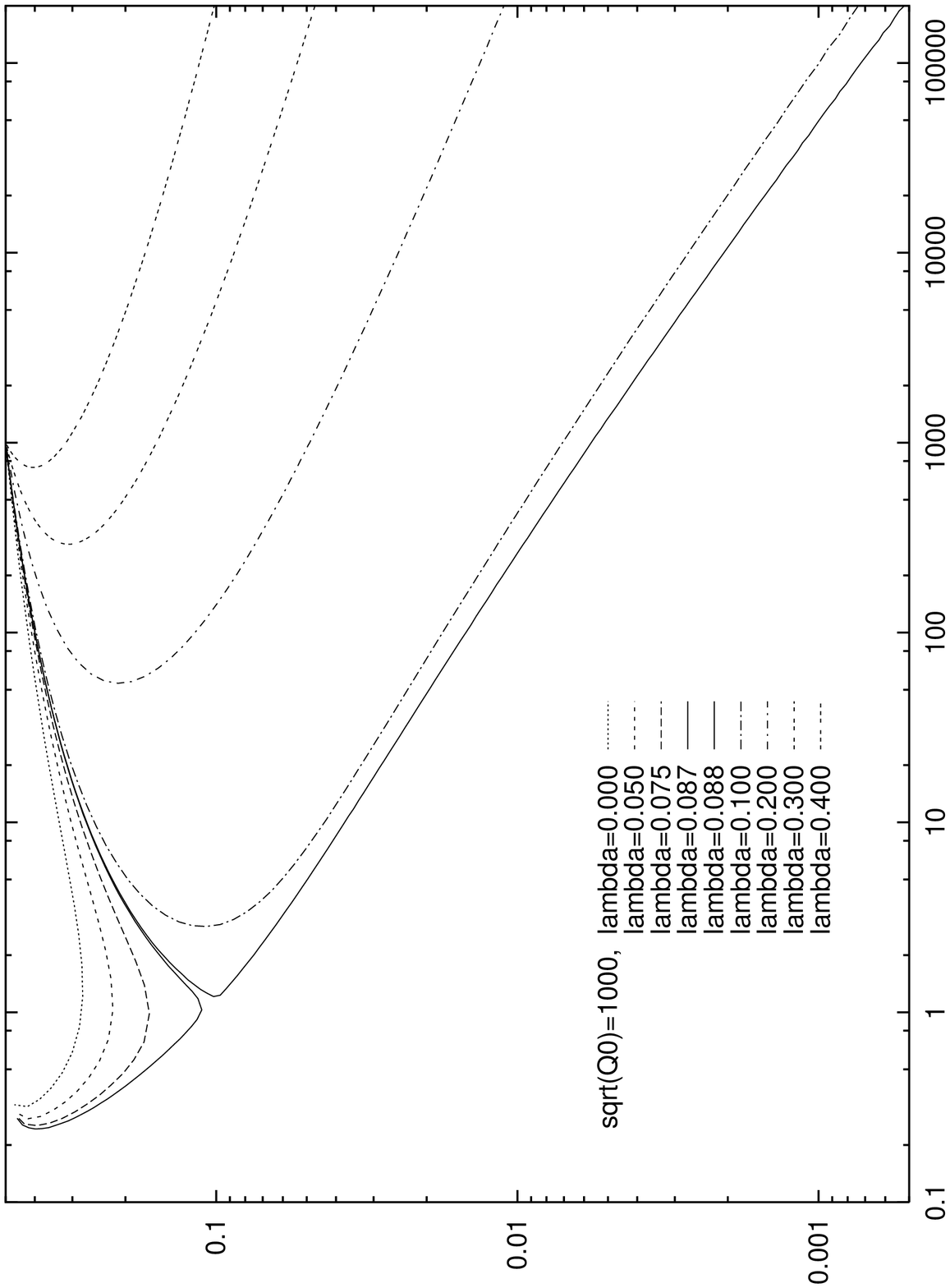} 
\includegraphics{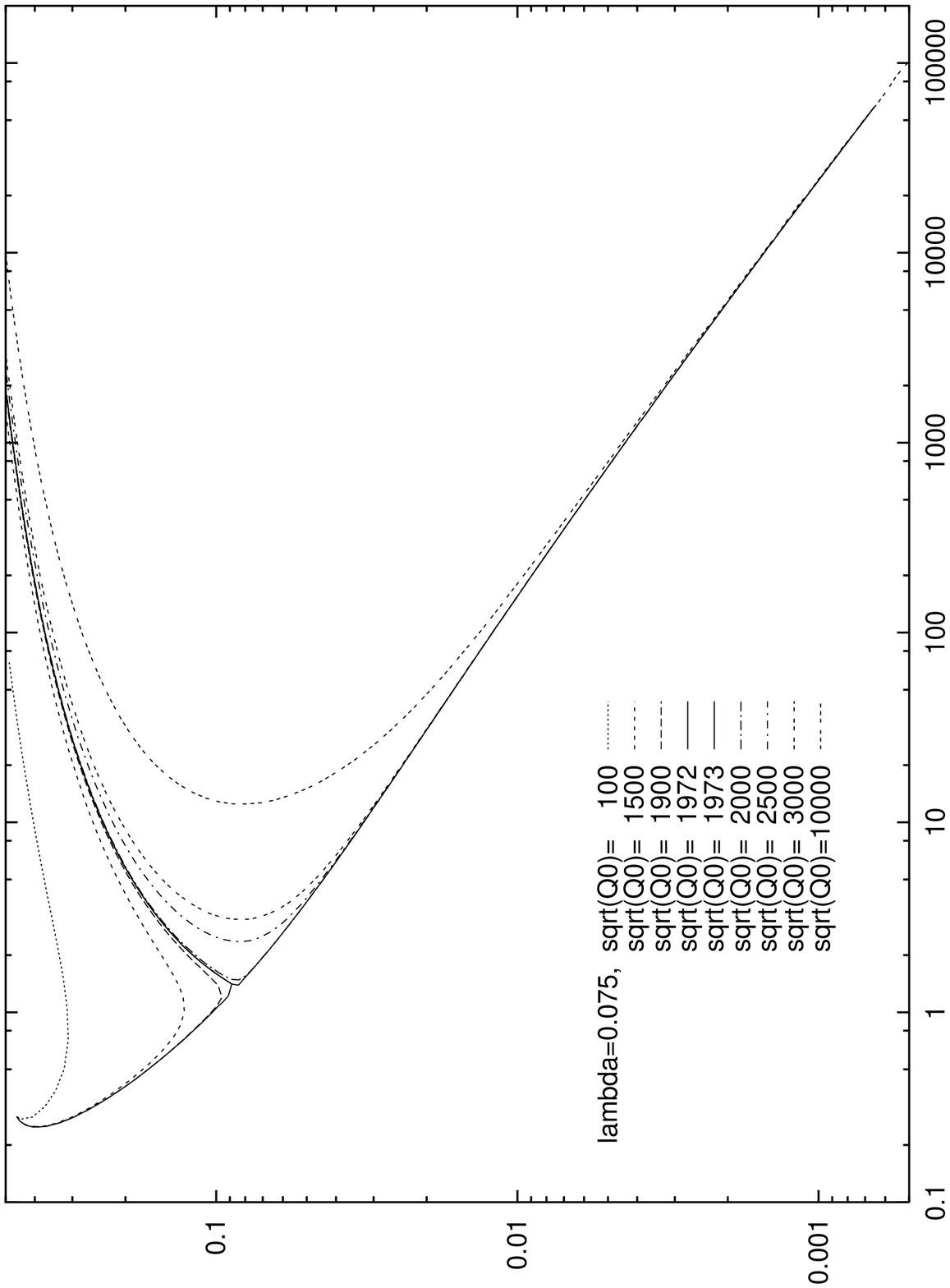} 
\caption{{\it Evolution of the generalization error $\epsilon_g$ (vertical 
axis) and of $\sqrt{{\cal Q}}$ (horizontal axis) at $L=10$ for various 
$\lambda$ and starting points ${\cal Q}(0)$}. 
} 
\label{f.d468} 
\end{figure}

To establish the asymptotic behaviour we look for solutions of 
the equations (\ref{e.deps},\ref{e.dqel}) in the limit of small $\epsilon_g$, 
large ${\cal Q}$. To leading order (for $\lambda > 0$), these equations become: 
 
\bea 
\frac{d\epsilon_g}{d\alpha} &\simeq& -\frac{\epsilon_g}{\sqrt{2\pi}L\sqrt{\cal Q}} 
+ \frac{\lambda^2}{2\pi^2 {\cal Q}\epsilon_g}, \\ 
\frac{d \sqrt{\cal Q}}{d\alpha} &\simeq&\sqrt{\frac{2}{\pi}}\lambda, 
\eea 
 
\no which can be solved exactly to give: 
 
\bea 
\epsilon_g^2 &\simeq& \frac{\lambda}{\sqrt{2\pi}\pi(\frac{1}{\lambda L}-1)} 
{\cal Q}^{-1/2} + c_1 {\cal Q}^{-\frac{1}{2\lambda L}}\ \ \ \ \ 
{\rm for}\  
\lambda \neq \frac{1}{L}, \label{e.eas1}\\ 
\epsilon_g^2 &\simeq&\left(\frac{1}{\pi\sqrt{2\pi}L} {\rm ln}{\cal Q}^{1/2} + c_2\right) 
{\cal Q}^{-1/2} \ \ \ \ \ {\rm for}\ \lambda = \frac{1}{L},  \label{e.eas2} 
\eea 
 
\no  i.e. explicitly
 
\bea
\epsilon_g^2 &\simeq& \frac{1}{2\pi(\frac{1}{\lambda L}-1)} 
{\alpha}^{-1} + {\tilde c}_1 {\alpha}^{-\frac{1}{\lambda L}}\ \ \ \  \ 
{\rm for}\  
\lambda \neq \frac{1}{L}, \label{e.easa1}\\ 
\epsilon_g^2 &\simeq&\left(\frac{1}{2\pi} {\rm ln}{\alpha} + 
{\tilde c}_2\right) 
{\alpha}^{-1} \ \ \ \ \ 
{\rm for}\ \lambda = \frac{1}{L},  \label{e.easa2}\\ 
 {\cal Q} &\simeq& \frac{2}{\pi}\lambda^2 \alpha^2 \label{e.qas} 
\eea 
 
\no asymptotically at large $\alpha$.

We see that for $\lambda < \frac{1}{L}$ we obtain asymptotically 
perfect generalization, 
the dominant term exhibiting  
 the usual power -1/2 (and, for $L=1,\ \lambda=0.5$, also the usual  
coefficient \cite{val}), while for $\lambda > \frac{1}{L}$ the second term in 
(\ref{e.eas1},\ref{e.easa1}) dominates and   ensures again 
 perfect generalization but with a different power law, $-1/(2\lambda L)$. 
For $\lambda = \frac{1}{L}$ we obtain logarithmic corrections -- see eq.  
(\ref{e.eas2},\ref{e.easa2}). Notice that these results hold also for $L=1$.

In the case $\lambda = 0$ one can see from (\ref{e.deps},\ref{e.dqel})
 that starting with any finite ${\cal Q}$ one cannot have perfect 
generalization for $L > 1$. For $L=1$ one reobtains the asymptotic behavior found in 
\cite{biri}.

\begin{figure}[htb] 
\vspace{7cm} 
\includegraphics{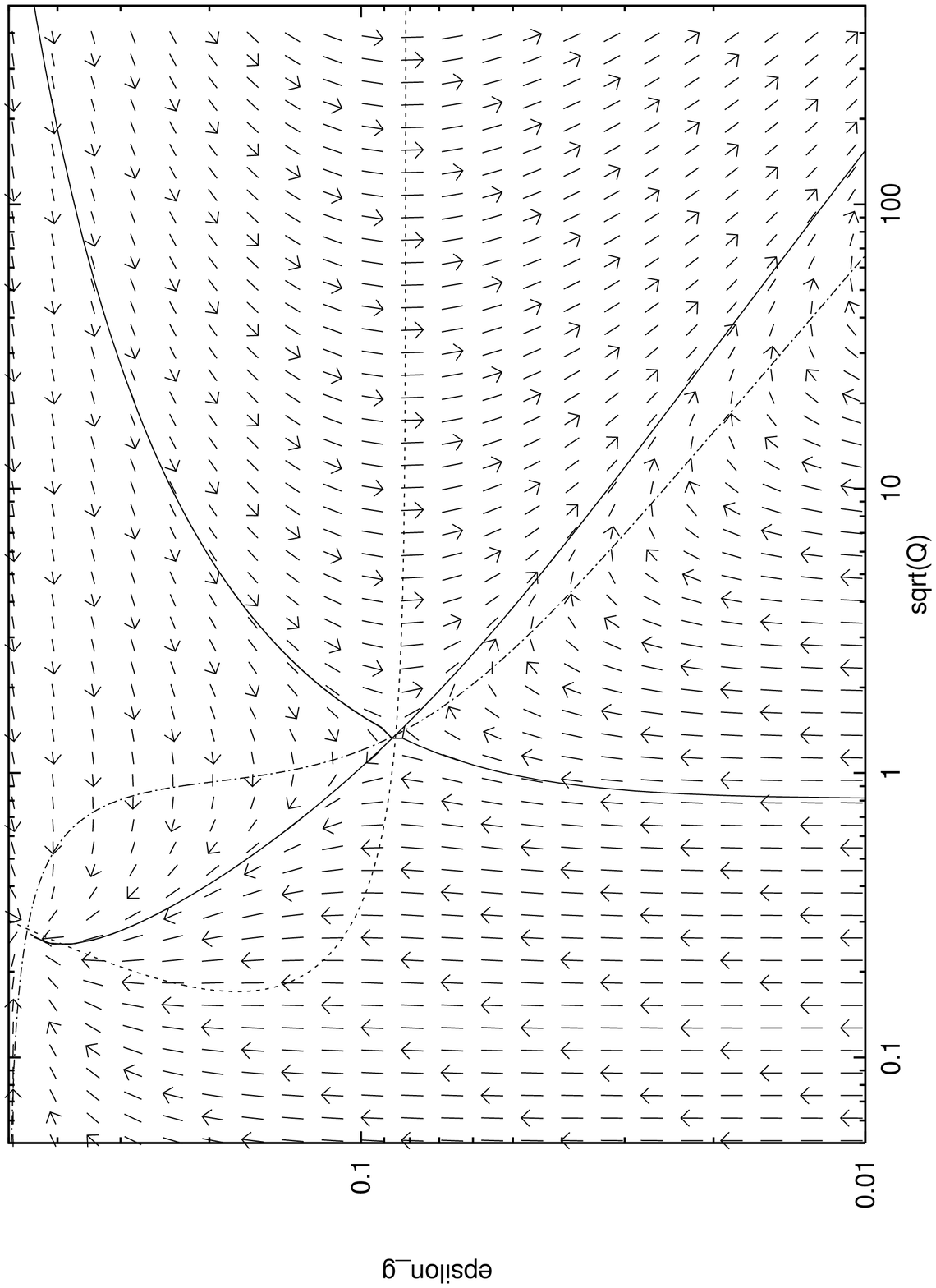} 
\includegraphics{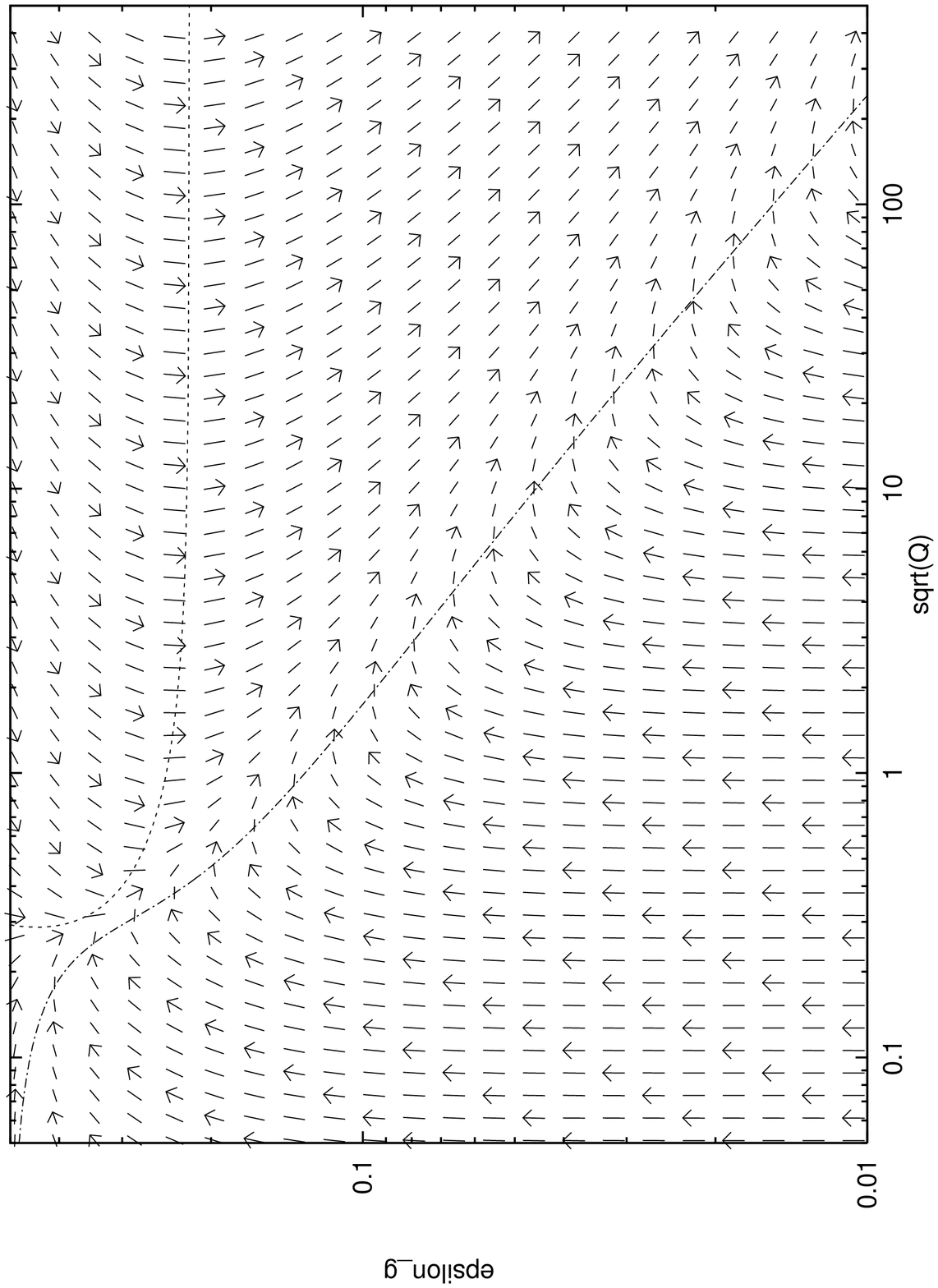}
\caption{{\it Flow in the plane $\epsilon_g$, ${\cal Q}$. The dot-dashed line corresponds 
to stationarity condition $d\epsilon_g/d \alpha = 0$, the dashed 
line to $d Q /d \alpha = 0$. Left: $L=10$, $\lambda=0.075$; two 
fixed points (one all stable, one partially stable) clearly show up, the full lines 
represent the stable and the unstable manifolds of 
the partially stable fixed point. Right:  
$L=5$, $\lambda=0.2$, a parameter setting for which there are 
no fixed points. For every starting point we 
have convergence to perfect generalization. Compare also with Fig. \ref{f.d468}}.
} 
\label{f.fp} 
\end{figure} 
 
There is, however, a nontrivial pre-asymptotic region, which turns out to 
be dominated by two stationarity conditions, one for the self-overlap, $d {\cal 
Q}/d \alpha = 0$, and one for the overlap with 
the teacher-configuration, 
$d {\cal R}/ d\alpha = 0$ or, alternatively, that for the generalization error
$d\epsilon_g/d \alpha = 0$. For suitable values of the network parameters,
the two stationarity conditions may simultaneously be satisfied, leading to
fixed points of the learning dynamics, one of these fully stable and 
with poor generalization, the other partially stable.

To this pre-asymptotic region we shall now turn our attention and thereby 
also obtain further specifications for the parameters.  
In Fig. \ref{f.d468} we show the evolution of $\epsilon_g$ and ${\cal Q}$ 
according to eqs. (\ref{e.deps},\ref{e.dqel}), starting from  
$\epsilon_g(0)=0.5$ and various ${\cal Q}(0)= {\cal Q}_0$.\footnote{Notice 
that due (\ref{e.eno}) the dependence on the initial conditions 
${\cal Q}_0$ may be translated into a dependence on the learning rate
for the initial network: for a fixed ratio $\lambda$ of learning rates,
and given  values of the original overlaps $Q$ and $R$, finer updating 
(smaller $a_1$ and $a_2$) is equivalent to larger values of rescaled 
overlaps, hence a larger value of ${\cal Q}_0$.}
The various trajectories are parameterized by $\lambda$. 
In all cases there is a critical value $\lambda_c({\cal Q}_0)$ 
which separates flows toward a stationary state
of poor generalization from flows toward perfect asymptotic generalization. 
The fixed point in the ${\cal Q},~ \epsilon_g$ plane 
(with a location parameterized by $\lambda$) which is responsible for this behaviour has an 
attractive and a repulsive direction. For a given initial condition ${\cal 
Q}_0$, the critical value $\lambda_c({\cal Q}_0)$ is defined as that value 
for which the attractive manifold connects the initial condition to the 
partially stable fixed point; for smaller values of $\lambda$ the flow always 
is from the initial condition to the fully stable fixed point with poor
generalization, for slightly larger values of $\lambda$ the flow is towards 
asymptotically perfect generalization. At still larger values of $\lambda$ 
the two fixed points eventually coalesce and disappear altogether. Then we 
always have asymptotically perfect generalization. Some values for 
$\lambda_c({\cal Q}_0)$ are given in Table \ref{t.tlc}.

\begin{table}[ht]
\begin{center}
\begin{tabular}{|c|c|c|c|c|c|}
\hline
 $\sqrt{{\cal Q}_0}$&1 &10& 100& 1000& 10000 \\ \hline 
$\lambda_c({\cal Q}_0)$& 0.2545(5)&0.2185(5)&0.1385(5)& 0.0875(5)& 0.0485(5)\\ \hline
\end{tabular}
\caption[]{{\it Critical value of $\lambda$ for $L=10$ and various initial conditions.}}
\label{t.tlc}
\end{center}
\end{table}

\begin{figure}[htb] 
\vspace{8.5cm} 
\includegraphics{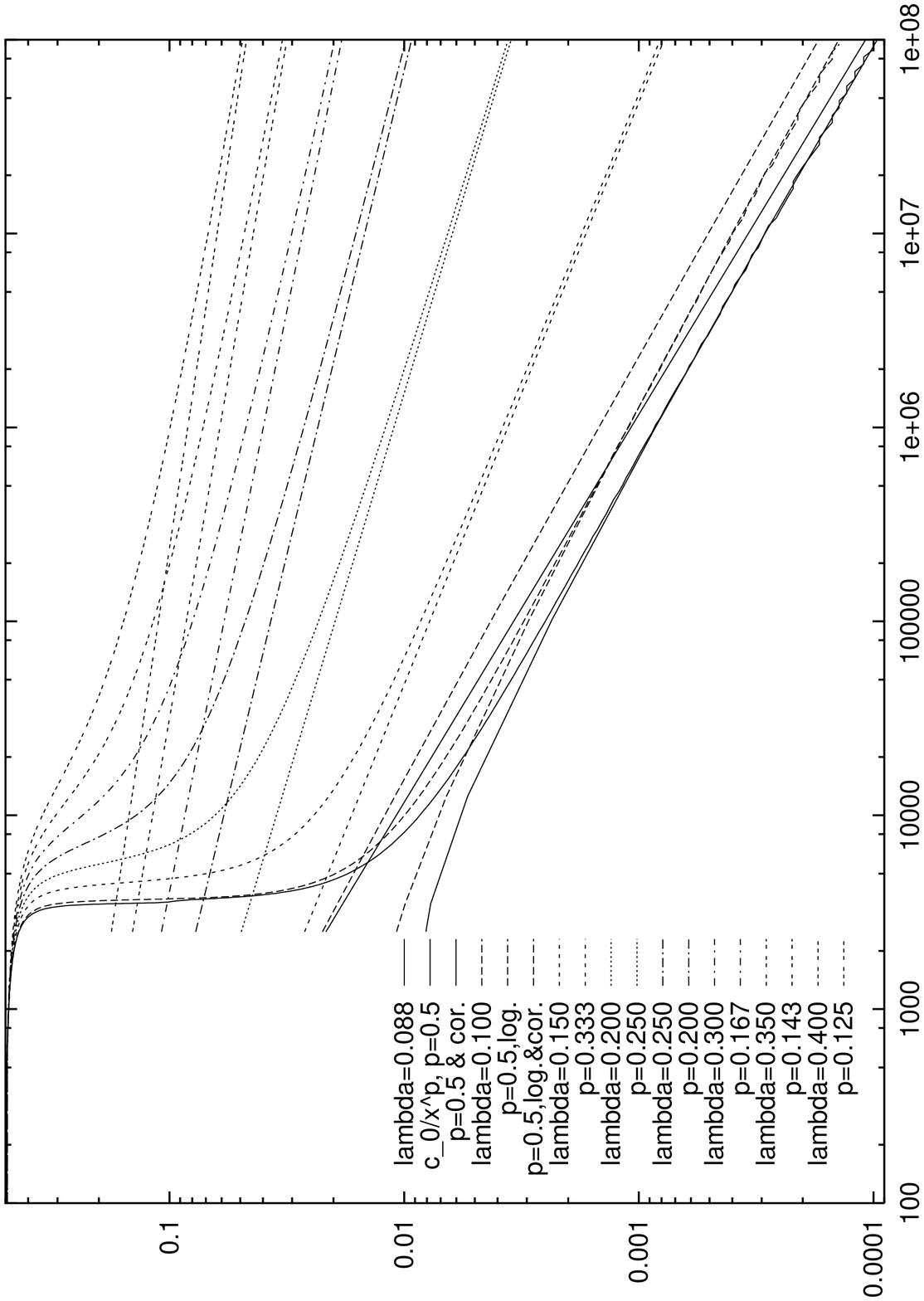} 
\includegraphics{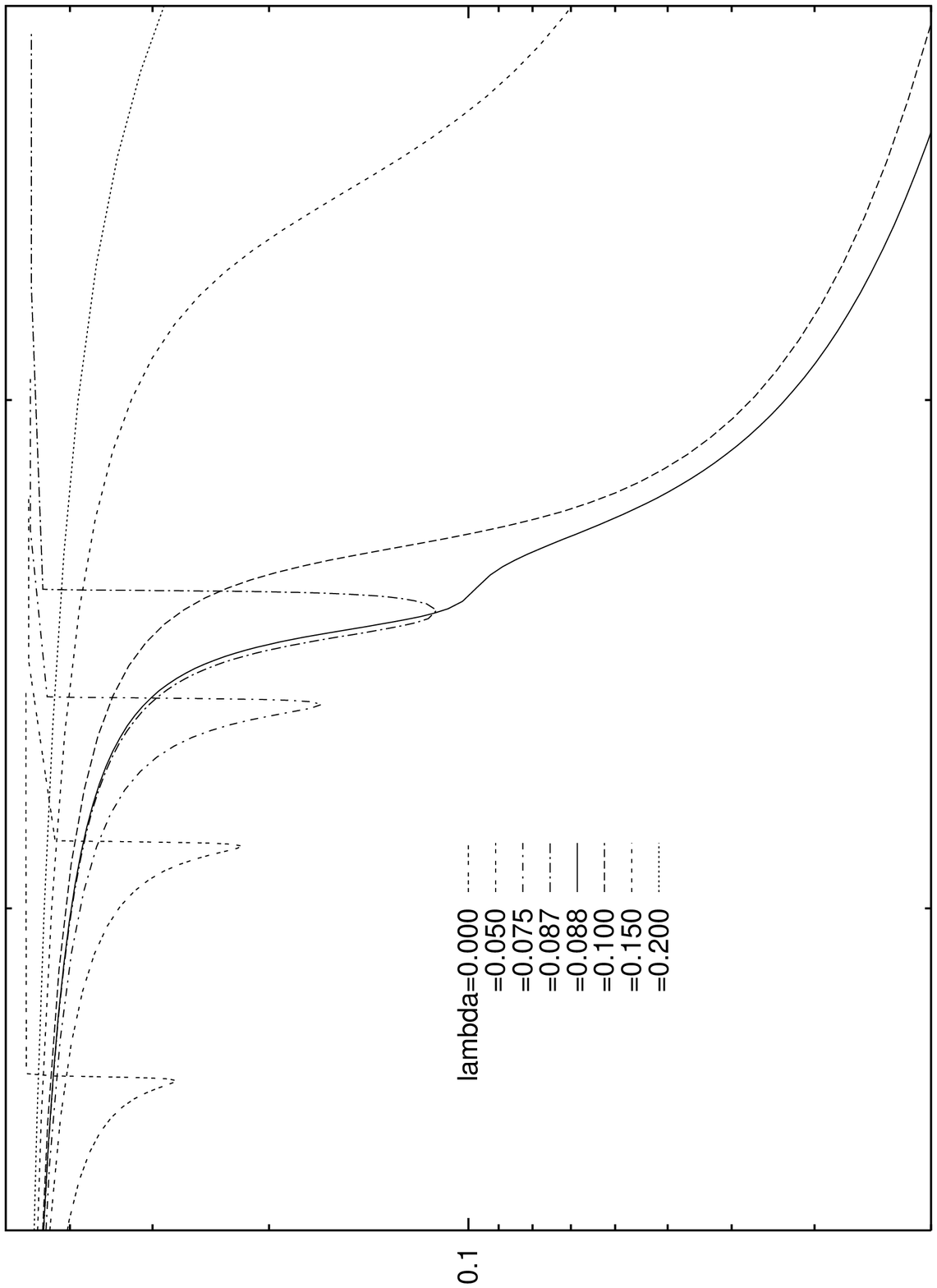} 
\caption{{\it Generalization error $\epsilon_g$ vs. $\alpha$ at $L=10$ for 
various $\lambda$ and for starting point ${\cal Q}(0)=1000$. The straight 
lines on the left plot show the dominant asymptotic behaviour for the 
corresponding $\lambda$ from (\ref{e.easa1},\ref{e.easa2}) (notice that
for  $\lambda \le 1/L$ the normalization is fixed; for $\lambda \le 1/L$ 
we give also a fit using the subdominant terms in 
(\ref{e.easa1},\ref{e.easa2})).
 Right: Amplified view at the pre-asymptotic 
region.}}
\label{f.a6} 
\end{figure}

In Fig. \ref{f.fp} we describe the flow in this plane for a given 
$\lambda$, this  should be compared with the  $\alpha$-trajectories in the  
${\cal Q},~ \epsilon_g$ plane  for various  $\lambda$ with different 
starting points ${\cal Q}_0$, Fig. \ref{f.d468}. In Fig. \ref{f.a6} we 
plot directly $\epsilon_g(\alpha)$. As can be seen from all these figures, 
for $\lambda < \lambda_c$ the training  leads to an initial improvement 
which is however limited and followed by a very rapid deterioration  
toward confusion. For $\lambda > \lambda_c$, on the contrary, the  
learning stabilizes and leads to asymptotically perfect generalization 
with a $\lambda$-dependent power law in agreement with eqs. 
(\ref{e.easa1},\ref{e.easa2}).
 
These analytic results compare very well with the numerical results 
given in the previous section, both in the pre-asymptotic and in the 
asymptotic region (cf. Fig. \ref{f.sim}). 
 
\section{Summary and Discussion} 

In the present paper we have investigated a two-phase learning algorithm
for perceptrons, named AR-Hebb-algorithm. 
Its first phase consists of a series of Hebb-type
synaptic modifications, correlating, however, input and {\em self-computed\/} 
output (blind association) rather than input and clamped teacher output. 
This first phase is followed by an unspecific  
but graded reinforcement-type learning step 
which leads to a partial reversal of the previous series of 
Hebb--type synaptic modifications, depending on current average success rates.

Our main motivation has been biological, attempting to honour the observation
that a learner's control over its neurons and synapses might be less specific
and direct than ordinary supervised learning algorithms usually presume, while
basically adhering to the Hebbian learning paradigm.

Our central results can be stated as follows:\par 
\no a) Despite the fact that feedback on the
learner's performance enters its learning dynamics only in an {\em 
unspecific\/} way in that it cannot be associated with single identifiable
correct or incorrect associations, convergence of the AR-Hebb-algorithm  in 
the sense of {\em asymptotically perfect generalization\/} is observed.\par

\no b) For given initial conditions, this convergence depends on the  
parameters of the algorithm; in particular none of these parameters can be 
set to 0. Alternatively, at fixed $L$ and ratio of the
algorithm parameters convergence may depend 
on initial conditions. 

In the details the dynamics of this algorithm was found to be unexpectedly complex. 
Depending on the parameters, fixed points in the dynamic flow may emerge ---
one stable, the other only partially stable. The attracting manifold of the
latter constitutes a separatrix dividing initial states into two sets, one
for which the algorithm converges, and another for which it doesn't 
in which case the 
flow is driven to the all--stable fixed point with poor generalization. 
Seen from a different point of view, a {\em given initial condition\/} 
(given updating speed) may 
be found to belong to the asymptotically converging lot, or to end up in 
a state of poor generalization, depending on network parameters. 

On the
other hand, parameter settings may be varied in such a way that the two fixed
points eventually coalesce and disappear, rendering convergence of the 
algorithm independent of initial conditions. The pre-asymptotic regime of
the learning process is then still influenced by the lines
in the $\epsilon_g$--${\cal Q}$ plane along which either $d\epsilon_g/d \alpha$
or $d {\cal Q}/d \alpha$ (but not both) vanish.

Much to our surprise, the {\em convergence--rate\/} of the algorithm was found
to depend in a {\em non-universal manner\/} on the ratio of learning
parameters. In spite 
of the non-specific nature of the information--feedback on the learning 
dynamics, convergence can be as fast as that of Hebbian learning, $\epsilon_g
\sim \alpha^{-1/2}$, if $\lambda L < 1$, whereas it is slower and exhibits 
a non-universal parameter dependent rate, $\epsilon_g \sim \alpha^{-1/2 
\lambda L}$, if $\lambda L > 1$. Logarithmic corrections appear in the 
marginal case $\lambda L = 1$.

One may ask oneself, why there is no generalization for a 
perceptron--type algorithm $\lambda =0$ (i.e., $a_1=0$). 
We can offer a simple observation which may be of heuristic value: 
since for $L=1$ $e_q$ can only be 0 or 1 
$a_1=0$ means penalty for failure, no change for success, i.e. the
usual perceptron learning rule known to converge. However, for $L>1$
$e_q$ can take fractional values in the interval $[0,1]$. In this case
$a_1=0$ means penalty for all answers which are short of perfect, i.e.
even if the pupil is successful in far above $50\%$ of the cases. This
procedure can turn out to be destructive.

To put our findings into a broader perspective, it is perhaps appropriate to
note that a similar kind of unspecific information feedback as in our setup
occurs in committee--machine learning. While in our case, information
feedback is unspecific in time (with respect to the pattern labels within
a longer series on which the learner may have been in error), unspecificity
in the committee--machine refers to space, i.e., the label(s) of the node(s) 
which may have contributed to a wrong output upon presentation of a single
pattern. In the details, though, the way in which unspecific feedback is utilized in the dynamics is different in the two setups, leading to different
asymptotic laws, and to different behaviour in the preasymptotic regime.
Although plateaus in the learning dynamics occur in both setups, this 
similarity is superficial. Whereas in the committee-machine, the appearance 
of plateaus is related to a permutation symmetry of the nodes and escape 
therefrom to its breaking (a transition to specialization), there is strictly 
speaking no time-translation symmetry within a coarse-grained step, and no 
breaking thereof, as each coarse-grained step constitutes a whole correlated
path of events during which the learner already evolves in response to the patterns presented. Quantitatively the difference manifests itself in the
fact that plateaus in our setup have a much higher generalization error than
those in the committee-machines, and that the AR-Hebb rule may converge to
a state of poor generalization even if its its initial performance is 
almost perfect (as can be seen in Fig. \ref{f.fp}a). Still, it may be 
interesting to enquire whether techniques akin to those invented in order to 
decrease the extent of plateaus in committe-machine learning (see \cite{boe98}
for a recent reference) might be utilized to improve the present setup.

We have not up to now addressed issues related to optimal parameter settings
or optimal online-control of parameters (the latter issue would in some 
sense run against our original biologically minded starting point), nor did
we so far investigate the performance of the algorithm in multi--layer 
architectures. Clearly these may be interesting topics to pursue in future 
research, as may be more detailed investigations of the algorithm as an 
intricate dynamical system per se.

{\bf Note added in proof:} We should like to add the following interesting 
observation. A variant of the present algorithm which introduces an additional 
biologically motivated element of indeterminism by including patterns in the 
second (reinforcement) phase of a session only with probability $p < 1$ 
(the student does not remember everything it did in the first phase) shows 
qualitatively the same behaviour as the algorithm studied in the present 
paper. A rough first quantitative characterization of this modification would 
be that it leads to an effective rescaling of the parameter $a_2$ of the 
algorithm by approximately a factor $p$, entailing corresponding rescalings 
of the parameter $\lambda$ and the scaled self-overlap ${\cal Q}$, viz. 
$\lambda \to \lambda/p$ and ${\cal Q} \to {\cal Q}/p^2$. Asymptotic convergence 
will then be slower, with an exponent computed from the rescaled value 
$\lambda/p$ rather than from the bare $\lambda$. It also leads to a 
corresponding reduction of critical $\lambda$'s for given initial condition 
$Q_0$ or, alternatively, to a reduction of the minimum $Q_0$ required for 
convergence at a given bare $\lambda$. These results are well corroborated by 
numerical simulations.

\end{document}